\documentclass[aps,prd,twocolumn,groupedaddress,showpacs]{revtex4}

\usepackage{graphicx}
\usepackage{dcolumn}

\newcommand{\bd}{\begin{displaymath}}
\newcommand{\ed}{\end{displaymath}}
\newcommand{\be}{\begin{equation}}
\newcommand{\ee}{\end{equation}}
\newcommand{\bda}{\begin{eqnarray*}} 
\newcommand{\eda}{\end{eqnarray*}}
\newcommand{\bea}{\begin{eqnarray}} 
\newcommand{\eea}{\end{eqnarray}}

\newcommand{\ms}{\overline{\mathrm{MS}}}
\def\slash#1{\mkern-1.5mu\raise0.4pt\hbox{$\not$}\mkern1.2mu #1\mkern 0.7mu}

\begin{document}

\title{Perturbative renormalization of the first moment of structure functions 
       \\ for domain-wall QCD}

\author{Stefano Capitani}
\email[]{stefano.capitani@uni-graz.at}
\affiliation{Institut f\"ur Physik, FB Theoretische Physik \\
Universit\"at Graz, A-8010 Graz, Austria}

\begin{abstract}
Using the domain-wall formulation of lattice fermions, we have computed the 
one-loop renormalization factors of one-link operators which measure the first 
nontrivial moment of the unpolarized, polarized and transversity structure 
functions, in the flavor nonsinglet sector. The knowledge of these factors is 
necessary in order to extract physical numbers from domain-wall Monte Carlo 
simulations of parton distributions.

We have automated the perturbative calculations by developing suitable FORM 
codes. The results show that in many instances the total renormalization 
factors are almost equal to one, and that hence the corresponding operators 
are, for the appropriate values of the Dirac mass $M$ and the coupling $g_0$, 
practically unrenormalized.
\end{abstract}

\pacs{12.38.Gc,11.10.Gh,13.60.Hb}

\maketitle

\section{Introduction}

Domain-wall fermions \cite{Kaplan:1992bt,Shamir:1993zy,Furman:1994ky}
provide a solution of the Ginsparg-Wilson relation \cite{Ginsparg:1981bj},
and as such they possess an exact chiral symmetry at nonvanishing lattice
spacings \cite{Luscher:1998pq} without at the same time presenting 
inconvenient features like doublers or nonanaliticities. They constitute 
one of the most promising formulations for simulations of chiral fermions 
on a lattice and for the study of physical issues connected with chirality 
\cite{Kikukawa:2001mw}. Although Monte Carlo simulations of these fermions 
require more computational efforts compared with some other nonchiral 
formulations (like Wilson fermions), recently many advances have been 
reported and at present domain-wall fermions are widely used in a variety 
of physical situations, for which some of the most recent results and 
investigations can be found in \cite{Aoki:2005ga,Dawson:2005zv,Edwards:2005an,Hashimoto:2005re,Lin:2005cn,Noaki:2005zw,Yamada:2005dv}.

Form factors, structure functions and generalized parton distributions 
are also among the phenomenological quantities which have been studied 
by means of simulations with domain-wall fermions 
\cite{Bistrovic:2005za,Edwards:2005kw,Edwards:2005ym,Orginos:2005uy}.
The calculation of the perturbative renormalization of the operators related 
to the moments of the deep inelastic structure functions, involving the 
treatment of covariant derivatives, has been missing up to now. These 
renormalization factors, whether perturbatively or nonperturbatively computed, 
are however necessary for the reliable extractions of physical numbers from 
Monte Carlo simulations of structure functions. The intention of the present 
work is to provide some of these factors from perturbation theory, and we have 
here considered the lowest nontrivial moment of various parton distributions.
In particular, we present results for the momentum, helicity and transversity
distributions, which give a complete description of the quark momentum and spin
at leading twist. 

This article is organized as follows. In Sect. \ref{sec:pt} we review the basic
perturbative ingredients which are necessary for the calculations presented 
in this work, and in Sect. \ref{sec:ren} we discuss the peculiar aspects of 
one-loop renormalization for domain-wall fermions. In Sect. \ref{sec:op} then
we introduce the operators of which we have evaluated the renormalization 
factors, which we give in Sect. \ref{sec:res}. Finally, in Sect. 
\ref{sec:concl} we make some concluding remarks, and in the Appendix we give 
the results for the quark self-energy and the bilinear operators, where we 
have found some discrepancies with old calculations \cite{Aoki:1998vv}, 
which however derive only from some constants in divergent continuum integrals.
These discrepancies do not affect the expressions of the renormalization 
factors, for which there is complete agreement with Ref. \cite{Aoki:1998vv}.

\section{Perturbative domain-wall}
\label{sec:pt}

We employ the standard formulation of domain-wall fermions devised by Shamir 
\cite{Shamir:1993zy}, where the five-dimensional quark action is given by
\bea
S^{DW}_q &=& \sum_x \sum_{s=1}^{N_s} \Bigg[ 
\frac{1}{2} \sum_\mu \Big( \overline{\psi}_s(x)
(\gamma_\mu - r) U_\mu (x) \psi_s(x+\hat{\mu}) 
\nonumber \\
&& - \overline{\psi}_s(x)
(\gamma_\mu + r) U^\dagger_\mu (x-\hat{\mu}) \psi_s(x-\hat{\mu}) \Big) 
\nonumber \\
&& + \Big( \overline{\psi}_s(x) P_+ \psi_{s+1}(x)
         + \overline{\psi}_s(x) P_- \psi_{s-1}(x) \Big)
\nonumber \\
&& + (M -1 +4r) \overline{\psi}_s(x) \psi_s(x) \Bigg] \\
&& + m \sum_x \Big( \overline{\psi}_{N_s}(x) P_+ \psi_1(x) 
+ \overline{\psi}_1(x) P_- \psi_{N_s}(x) \Big) . \nonumber 
\label{eq:dwaction}
\eea
The Wilson parameter is set to $r=-1$, and the Dirac mass $M$ takes values 
between zero and two (at tree level) so that the correct structure of chiral 
modes (with no doublers) is attained for $N_s\rightarrow\infty$. The chiral 
projectors are $P_\pm = (1 \pm \gamma_5)/2$. Here and in most of the paper 
we put $a=1$. 

The above domain-wall action can be imagined as a Wilson action endowed with 
an additional flavor index $s$ plus a special mass matrix for these flavors, 
explicitly given in Eqs.~(\ref{eq:massmatrix1}) to (\ref{eq:massmatrix4}) 
further below. The mass matrix governs the mixing among the flavors and induces
a sophisticated structure on the flavor space, which at the end produces one 
light quark and $N_s-1$ heavy quarks. For this reason the tree-level 
quark propagator turns out to have a more complicated form than in the 
four-dimensional Wilson case, and in practical terms perturbation theory 
for domain-wall fermions looks like having $N_s$ fermion flavors with an 
involved propagator structure in the index $s$. The gluon fields and their 
couplings to the quarks are instead kept four-dimensional, that is they 
do not depend on the fifth dimension and are identical at each $s$. 
The gluon propagator and vertices are then just the same as in a 
four-dimensional lattice theory. In this work we have used for the pure gauge 
part of the domain-wall action the standard plaquette action, and we perform 
all computations in a general covariant gauge, where the gluon propagator 
is given by
\be
G_{\mu\nu}(k) = \frac{1}{4\sum_\rho \sin^2 \frac{k_\rho}{2}}
\Bigg( \delta_{\mu\nu} - (1-\alpha) \frac{4 \sin \frac{k_\mu}{2} 
\sin \frac{k_\nu}{2}}{4\sum_\lambda \sin^2 \frac{k_\lambda}{2}} \Bigg) ,
\ee 
where $\alpha=1$ and $\alpha=0$ correspond to the Feynman and Landau gauges 
respectively. The measure term, the gauge-fixing term and the Faddeev-Popov 
term, as well as the quark-gluon interaction vertices, have also the same 
expression as in the Wilson case. Since $r=-1$, the vertices that we need 
in this paper assume the form (apart from color factors)
\bea
V^{(1)}_\mu (p) &=& - g_0 \, \Big( i \gamma_\mu \cos \frac{p_\mu}{2}
              - \sin \frac{p_\mu}{2} \Big)  \\
V^{(2)}_{\mu\nu} (p) &=& \frac{1}{2} \, g_0^2 \, \Big( i \gamma_\mu 
   \sin \frac{p_\mu}{2} + \cos \frac{p_\mu}{2} \Big) \cdot \delta_{\mu\nu}
\eea
for the interaction of the quark current with one gluon and two gluons 
respectively, where $p$ stands for the sum of the incoming and outgoing 
quark momenta.

The construction of the tree-level quark propagator has been started in 
\cite{Narayanan:1992wx,Shamir:1993zy} and 
\cite{Aoki:1993rg,Aoki:1996zz,Vranas:1997da,Kikukawa:1997tf} and then completed
and used in in the first full-fledged calculations of the renormalization of 
the quark self-energy and bilinears \cite{Aoki:1997xg,Aoki:1998vv,Blum:1999xi}.
Further perturbative results for domain-wall fermions have been obtained in 
\cite{Aoki:1998hi,Aoki:1999ky,Noaki:1999ij,Aoki:2000ee,Aoki:2002iq,Aoki:2003uf,Yamada:2004ri}, and some perturbative calculations for nonstandard domain-wall
actions have been carried out in \cite{Shamir:2000cf}.

For our perturbative calculations with domain-wall fermions we use the same 
framework of \cite{Aoki:1998vv,Blum:1999xi}, where one works solely with the 
dominant contributions to the propagators when the number of flavors $N_s$ 
(or points in the fifth additional dimension) goes to infinity. In this limit 
the decoupling of the chiral modes is exact for massless quarks, chirality 
breaking terms being exponentially suppressed in the length of the fifth 
dimension.

The domain-wall Dirac operator after a Fourier transform in (four-dimensional) 
momentum space becomes
\bea
D_{st} (p) &=& \delta_{s,t} \, \sum_\mu i \gamma_\mu \sin p_\mu \\
&+& (W^+_{st} (p) + mM^+_{st}) P_+ + (W^-_{st} (p) +mM^-_{st}) P_- , \nonumber
\eea
where the mass matrix is given by 
\bea
W^\pm_{st}(p) &=& -W(p) \, \delta_{s,t} + \delta_{s\pm 1,t} , \\
M^+_{st} &=& \delta_{s,N_s} \, \delta_{t,1} , \\
M^-_{st} &=& \delta_{s,1} \, \delta_{t,N_s} , 
\eea
and
\be
W(p) = 1 - M +2 \sum_\lambda \sin^2 \frac{p_\lambda}{2} .
\ee
In more explicit form,
\bea
W^+(p) &=& \pmatrix{
-W(p) & 1      &        &       \cr
      & -W(p)  & \ddots &       \cr
      &        & \ddots & 1     \cr
      &        &        & -W(p) \cr} ,
\label{eq:massmatrix1} \\
W^-(p) &=& \pmatrix{
-W(p) &        &        &       \cr
 1    & -W(p)  &        &       \cr
      & \ddots & \ddots &       \cr
      &        & 1      & -W(p) \cr} ,
\label{eq:massmatrix2} \\
M^+ &=& \pmatrix{
  &  &   \cr
  &  &   \cr
1 &  &   \cr} ,
\label{eq:massmatrix3} \\
M^- &=& \pmatrix{
  &  & 1 \cr
  &  &   \cr
  &  &   \cr} .
\label{eq:massmatrix4} 
\eea
We see that for $m=0$ the $M^\pm$'s are absent from the action and hence the 
propagator no longer possesses any terms directly connecting the two 
boundaries at $s=1$ and $s=N_s$.

In this work we only consider massless quarks. By inverting the above Dirac 
operator with $m=0$ one obtains the tree-level quark propagator
\begin{widetext}
\be
\langle \psi_s (-p) \overline\psi_t (p)\rangle
 = \sum_u \Big[ \Big( - i \gamma_\mu \sin p_\mu \,\delta_{s,u} 
                      +W^-_{su}(p)\,\Big)\, G^R_{ut}(p) \,P_+ 
 + \Big( - i \gamma_\mu \sin p_\mu \,\delta_{s,u} 
                      +W^+_{su}(p)\,\Big)\, G^L_{ut}(p) \,P_- \Big] .
\label{eq:psipsi}
\ee
The expressions of the functions $G^R(p)$ and $G^L(p)$ are, for large $N_s$, 
\bea
G^R_{st}(p) &=& - \frac{A(p)}{F(p)} \, \Big( 
(1-W(p)e^{-\alpha(p)}) \, e^{-(2N_s-s-t)\alpha (p)} 
+ (1-W(p)e^{\alpha(p)}) \, e^{-(s+t)\alpha (p)} \Big) 
+ A(p) \, e^{-|s-t|\alpha(p)} , \\
G^L_{st}(p) &=& - \frac{A(p)}{F(p)} \, \Big(
(1-W(p)e^{\alpha(p)}) \, e^{-(2N_s-s-t+2)\alpha (p)} 
+ (1-W(p)e^{-\alpha(p)}) \, e^{-(s+t-2)\alpha (p)} \Big) 
+ A(p) \, e^{-|s-t|\alpha(p)} ,
\eea
\end{widetext}
where $\alpha(p)$ is defined by the positive solution of the equation 
\cite{Narayanan:1992wx,Shamir:1993zy}
\be
\cosh (\alpha(p)) = \frac{1+W^2(p)+\sum_\lambda \sin^2 p_\lambda}{2|W(p)|} ,
\ee
and
\bea
A(p) &=& \frac{1}{2W(p) \, \sinh (\alpha (p))} , \\
F(p) &=& 1-W(p) \, e^{\alpha(p)} .
\eea
These formulae are only valid for positive $W$, which is always the case
if $0 < M < 1$. When $W$ has a zero, $\alpha$ has a logarithmic singularity. 
For $1 < M < 2$, $W$ can become negative if the momentum is small enough. 
In this case the propagator is given by the above equations with the 
replacements 
\bea
W &\rightarrow& - |W|, \\
e^{\pm \alpha} &\rightarrow& -e^{\pm \alpha} ,
\eea
which imply that also $\sinh \alpha$ changes sign.

To study matrix elements of the chiral modes in perturbation theory, we need 
to diagonalize the mass matrix in the fifth dimension. However, since this 
matrix is not hermitian, one has rather to consider the squared mass matrix, 
that is the second-order operators $D D^\dagger$ and $D^\dagger D$, which are 
hermitian and nonnegative and give a well-behaved spectrum. 
In the second-order operators the two chiralities are in fact well decoupled.

The chiral mode is obtained by means of a rotation in the fifth dimension of 
the original quark fields $\psi_s(x)$ to the basis which diagonalizes the mass 
matrix, and is given by
\be
\chi_0 (x) = \sqrt{1-w_0^2} \sum_s (P_+ w_0^{s-1} \psi_s (x) 
                            + P_- w_0^{N_s-s} \psi_s (x)) ,
\label{eq:chiralmode}
\ee
where
\be
w_0 = W(0) = 1 - M .
\ee
We can see that, because of the damping factors $w_0^{s-1}$ and $w_0^{N_s-s}$,
the chiral mode is exponentially localized near the boundaries of the fifth 
dimension. The ``physical'' quark fields that are instead used in Monte Carlo 
simulations are however somewhat simpler, and they are constructed only from 
quark fields exactly located at these boundaries:
\bea
q(x) &=& P_+ \psi_1 (x) + P_- \psi_{N_s} (x) \\
\overline q(x) &=& \overline\psi_{N_s} (x) P_+ + \overline\psi_1 (x) P_- .
\eea
We also use these these expressions for our calculations, as done in
\cite{Aoki:1997xg,Aoki:1998vv,Blum:1999xi}.

The computation of matrix elements involving states and operators constructed 
from these physical quark fields requires additional propagators. 
We need in fact to connect an internal with a physical quark field, 
and the corresponding propagators are given by
\begin{widetext}
\be
\langle q (-p) \overline\psi_s (p)\rangle =
\frac{1}{F(p)} \, i \gamma_\mu \sin p_\mu \,  
\Big( e^{-(N_s-s)\alpha(p)} P_+ + e^{-(s-1)\alpha(p)} P_- \Big)
\, - e^{-\alpha(p)} \Big( e^{-(s-1)\alpha(p)} P_+ + e^{-(N_s-s)\alpha(p)} 
P_- \Big) , 
\label{eq:qpsi} 
\ee
\be
\langle \psi_s (-p) \overline q (p)\rangle =
\frac{1}{F(p)} \, 
\Big( e^{-(N_s-s)\alpha(p)} P_- + e^{-(s-1)\alpha(p)} P_+ \Big) \, 
i \gamma_\mu \sin p_\mu
\, - e^{-\alpha(p)} \Big( e^{-(s-1)\alpha(p)} P_- + e^{-(N_s-s)\alpha(p)} 
P_+ \Big) .
\label{eq:psiq}
\ee
For our calculations it is also necessary to know their expansions for small 
momentum: 
\bea
\langle q (-p) \overline \psi_s (p)\rangle_c &=&
\frac{1-w_0^2}{i\slash{p}} \, \Big( w_0^{N_s-s} P_+ + w_0^{s-1} P_- - 
\frac{w_0}{1-w_0^2} \, i\slash{p} \, ( w_0^{s-1} P_+ + w_0^{N_s-s} P_- ) \Big)
\label{eq:qpsi0} , \\ 
\langle \psi_s (-p) \overline q (p)\rangle_c &=&
\Big( w_0^{N_s-s} P_- + w_0^{s-1} P_+ 
- ( w_0^{s-1} P_- + w_0^{N_s-s} P_+ ) \, \frac{w_0}{1-w_0^2}i\slash{p} \Big)
\, \frac{1-w_0^2}{i\slash{p}} ,
\label{eq:psiq0}
\eea
where the factors $1-w_0^2$ are related to the sums of the tree-level 
exponential damping factors over the fifth dimension:
\end{widetext}
\be
\lim_{N_s\rightarrow\infty} \, \sum_{s=1}^{N_s} \Big( w_0^{N_s-s} P_+ 
+ w_0^{s-1} P_- \Big)^2 = \frac{1}{1-w_0^2} .
\ee
Finally, we also need the tree-level propagator
\be
\langle q (-p) \overline q (p)\rangle = \frac{i \gamma_\mu \sin p_\mu}{F(p)} ,
\label{eq:qq}
\ee
which in the limit of small momentum is equal to
\be
\langle q (-p) \overline q (p)\rangle_c =
\frac{1-w_0^2}{i\slash{p}} .
\label{eq:qq0}
\ee

\section{One-loop renormalization}
\label{sec:ren}

Matrix elements estimated by importance sampling in Monte Carlo lattice 
simulations need to be properly renormalized in order to become meaningful 
physical numbers. They can be considered as (regulated) bare quantities, 
and in order to get physical results they have to undergo a lattice 
renormalization which matches them to some continuum scheme. We choose 
for the continuum the $\ms$ scheme of dimensional regularization, since 
commonly Wilson coefficients of operator product expansions are computed 
in this scheme.

A perturbative lattice renormalization involves both lattice and continuum 
perturbative calculations. At tree level, for momenta much lower than the 
lattice cutoff, lattice operators have the same matrix elements as the 
original continuum operators. At one loop one then gets, for the case of a
multiplicatively renormalized operator,
\be
\langle q | O^{lat} | q \rangle = \Big( 1 
+\bar g^2 \Big( -\gamma^{(0)} \log a^2p^2 
+ R^{lat} \Big) \Big) \cdot 
\langle q | O^{tree} | q \rangle  
\label{eq:1looplat} , 
\ee
\be
\langle q | O^{\ms} | q \rangle = \Big( 1 
+\bar g_{\ms}^2 \Big( -\gamma^{(0)} \log \frac{p^2}{\mu^2} 
+ R^{\ms} \Big) \Big) \cdot \langle q | O^{tree} | q \rangle ,
\label{eq:1loopcont}
\ee
where the lattice and continuum one-loop finite constants, $R^{lat}$ and 
$R^{\ms}$, do not have in general the same value, and hence the one-loop 
renormalization factors on the lattice and in the continuum are in general 
not equal (the one-loop anomalous dimensions are however the same). Here and 
in the following we call for brevity $\bar g^2 = (g_0^2/16 \pi^2)\, C_F$ 
(and similarly for $\bar g_{\ms}^2$), with $C_F=(N_c^2-1)/2N_c$ for the 
$SU(N_c)$ gauge group. 

The connection between the original lattice numbers and the final continuum 
physical results is given, neglecting higher-order terms in $\bar g^2$, by 
\cite{Martinelli:1982mw}
\be
\frac{\langle q | O^{\ms} | q \rangle}{\langle q | O^{lat} | q \rangle}
= 1-\bar g^2\Big( -\gamma^{(0)} \log a^2\mu^2 +R^{lat} -R^{\ms} \Big) ,
\label{eq:1loopcontlat}
\ee
where the difference $\Delta R = R^{lat} - R^{\ms}$ determines 
the renormalization factor
\be
Z_O (a\mu, \bar g)= 1 -\bar g^2 \Big( -\gamma^{(0)} \log a^2\mu^2 
+ \Delta R \Big)
\ee
which converts the lattice operator $O^{lat}$ into the physical renormalized 
operator $O^{\ms}$. The computation of these renormalization factors requires 
both lattice and continuum perturbative techniques (for more details see 
\cite{Capitani:2002mp}). In the domain-wall case it presents additional 
peculiar features that is worth reviewing. 

Let us first consider, in the massless case, the one-loop correction to the 
domain-wall quark propagator $\langle q (-p) \overline q (p)\rangle_c$.
It can be easily seen that, given the structure of the 
propagators $\langle q (-p) \overline\psi_s (p)\rangle_c$   
and $\langle \psi_s (-p) \overline q (p)\rangle_c$, we can write 
\bea
\langle q (-p) \overline q (p)\rangle_{1~loop} &=&
 \frac{1-w_0^2}{i\slash{p}} \, \Sigma_q (p) \, \frac{1-w_0^2}{i\slash{p}} \\ 
&=& \frac{1-w_0^2}{i\slash{p} - (1-w_0^2) \, \Sigma_q (p) } ,
\eea
where
\begin{widetext}
\bea
\Sigma_q (p) = & \sum_{s,t} & \Big( w_0^{N_s-s} P_+ + w_0^{s-1} P_- - 
\frac{w_0}{1-w_0^2} \, i\slash{p} \, ( w_0^{s-1} P_+ + w_0^{N_s-s} P_- ) \Big)
\nonumber \\
&& \cdot \Sigma_{st} (p) \\ 
&& \cdot \Big( w_0^{N_s-t} P_- + w_0^{t-1} P_+
- ( w_0^{t-1} P_- + w_0^{N_s-t} P_+ ) \, \frac{w_0}{1-w_0^2}i\slash{p} \Big) .
\nonumber 
\eea
\end{widetext}
The calculation of the one-loop self-energy diagrams gives \cite{Aoki:1998vv}
\be
\Sigma_{st} (p) = - \bar g^2 \, 
\Big( i\slash{p} ( I^+ P_+ + I^- P_- ) + W_1^+ P_+ + W_1^- P_- \Big)_{st},
\label{eq:sigmast}
\ee
and when the damping factors are also taken into account the final result 
can be written as
\be
\Sigma_q (p) =  \frac{1}{1-w_0^2} \, i\slash{p} \, \bar g^2 \, 
\Big( \alpha \log a^2 p^2 + \Sigma_1 - \frac{2w_0}{1-w_0^2} \, \Sigma_3 \Big) .
\label{eq:seint}
\ee

Putting all together, we see that the one-loop correction to the quark 
propagator is of the same form as its tree-level expression:
\be
\langle q (-p) \overline q (p)\rangle_{1~loop}
= \frac{1-w_0^2}{i\slash{p} - (1-w_0^2) \, \Sigma_q (p) } 
= \frac{1-w_0^2}{i\slash{p}} \, Z_w \, Z_2,
\ee
where 
\be
Z_2 =  1 + \bar g^2 \, \Big( \alpha \log a^2 p^2 + \Sigma_1 \Big) 
\ee
is the usual quark wave function renormalization factor, whereas
\be
Z_w =  1 - \frac{2w_0}{1-w_0^2} \, \bar g^2 \, \Sigma_3
=  1 + \bar g^2 \, z_w 
\ee
is a new feature appearing in domain-wall fermions, which represents an 
additive renormalization to $w_0$, as can be seen from
\be
(1-w_0^2) \, Z_w = 1 - \big( w_0 + \bar g^2 \, \Sigma_3 \big)^2 + O(\bar g^4).
\ee
Thus, while the zero mode remains stable under radiative corrections, 
the Dirac mass $M = 1 - w_0$ is additively renormalized. This effect is due 
to the $W_1^\pm$ terms in Eq.~(\ref{eq:sigmast}), which in turn originate 
from the order $a$ terms in the damping factors of Eqs.~(\ref{eq:qpsi0}) and 
(\ref{eq:psiq0}). We have by explicit calculation checked that the part 
proportional to $1-\alpha$ of $\Sigma_3$ is zero (its contribution from the 
half-circle diagram exactly canceling the one of the tadpole), which means 
that $\Sigma_3$ and $Z_w$ are gauge invariant. The values of $\Sigma_1$, 
$\Sigma_3$ and $z_w$ can be found in the Appendix.

We remark that in the above domain-wall self-energy there is no term 
$\Sigma_0$ proportional to $1/a$, which if present would signal a breaking 
of chirality.

Let us now consider a composite operator $\overline q(x) \, O \, q(x)$ which
is multiplicatively renormalizable. Again, by looking at the form of the 
propagators involved, one can see that the one-loop matrix element of this
operator between ``physical'' quark states is given by 
\be
\langle \, ( \, \overline q O q \, ) \, q \overline q \, \rangle_{1~loop} = 
\frac{1-w_0^2}{i\slash{p}} \cdot A_O (p) \cdot O \cdot 
\frac{1-w_0^2}{i\slash{p}} ,
\ee
where $A_O (p)$ contains the contribution of the damping factors and 
can be written as 
\be
A_O (p) = \bar g^2 \Big( - \gamma_O \log a^2 p^2  + B_O \Big).
\label{eq:matel1}
\ee
The one-loop expression has the same form as the tree-level matrix element. 
That also the self-energy contribution to the matrix element fits properly 
here can be seen (for example when $O=\gamma_\mu$) from inserting 
Eq.~(\ref{eq:seint}) (without $\Sigma_3$) in the expression of the contribution
of a leg in Fig.~\ref{fig:diagrams},
\be
\frac{1-w_0^2}{i\slash{p}} \cdot \frac{1}{1-w_0^2} \, i\slash{p}  
\cdot \bar g^2 \, \Big( \alpha \log a^2 p^2 + \Sigma_1 \Big) \cdot
\frac{1-w_0^2}{i\slash{p}} \, \gamma_\mu \,
\frac{1-w_0^2}{i\slash{p}} ,
\ee
which shows that indeed it gives a multiplicative correction to the 
tree-level matrix element:
\be
\frac{1-w_0^2}{i\slash{p}} \cdot \bar g^2 \, 
\Big( \alpha \log a^2 p^2 + \Sigma_1 \Big) \cdot \, \gamma_\mu \cdot
\frac{1-w_0^2}{i\slash{p}} .
\ee

\section{Structure function operators}
\label{sec:op}

The operators that we have considered in this work measure the lowest moment 
of various structure functions. They include all three parton distributions 
that characterize the quarks in the nucleon: the momentum distribution 
$q(x,Q^2)$ (described by the $F_1$ and $F_2$ unpolarized structure functions), 
the helicity distribution $\Delta q(x,Q^2)$ (described by the $g_1$ structure 
function), and the (chiral odd) transversity distribution $\delta q(x,Q^2)$ 
(described by the $h_1$ structure function). They thus provide a complete 
description of quark momentum and spin at leading twist.
We have also computed the renormalization of the lowest moment of the $g_2$ 
structure function, which receives contributions from twist-3 operators and 
measures the (chiral even) transverse spin. We refer for a more detailed 
discussion of these structure functions and in particular of the operators 
appearing in their operator product expansions, some of which are given below, 
to \cite{Capitani:2000wi,Capitani:2000aq} (of which we follow the notation) 
and references therein.

We have computed the renormalization factors of all flavor nonsinglet operators
which contain at most one covariant derivative. We have chosen in particular 
\bea
O_{v_2,d} &=& \bar q \gamma_{\{1} D_{4\}} q \label{eq:1ud} , \\
O_{v_2,e} &=& \bar q \gamma_4 D_4 q
               -\frac{1}{3} \sum_{i=1}^3  \bar q \gamma_i D_i q ,
\eea
which measure the first moment of the momentum distributions,
\bea
O_{a_2,d} &=& \bar q \gamma_{\{1} \gamma_5 D_{4\}} q \label{eq:1pd} , \\
O_{a_2,e} &=& \bar q \gamma_4 \gamma_5 D_4 q
            -\frac{1}{3} \sum_{i=1}^3  \bar q \gamma_i \gamma_5 D_i q ,
\eea
which measure the first moment of the helicity distributions,
\be
O_{d_1} = \bar q \gamma_{[4} \gamma_5 D_{1]} q ,
\ee
which taken together with $O_{a_2}$ determines the first moment of 
the $g_2$ structure function, and finally
\bea
O_{t_1} &=& \bar q \sigma_{41} \gamma_5 q , \\
O_{t_2} &=& \bar q \sigma_{4\{1} \gamma_5 D_{2\}} q ,
\eea
which correspond to the tensor charge and the lowest nontrivial moment of the
$h_1$ transversity structure function respectively. We have not explicitly 
shown the Gell-Mann flavor matrices which specialize them to nonsinglet 
operators and hence forbid any mixing with gluonic operators, because they are 
irrelevant for the sake of the calculation of the renormalization factors.
The symbol $\{\}$ denotes symmetrization over the relevant Lorentz indices,
while $[]$ denotes antisymmetrization. For the 
covariant derivatives $\stackrel{\phantom{\rightarrow}}{D}=
\stackrel{\rightarrow}{D}-\stackrel{\leftarrow}{D}$
we use the lattice discretizations
\be
\stackrel{\rightarrow}{D_\mu} q (x) = \frac{1}{2} \, \Big[U_\mu (x) 
q(x+\hat{\mu}) -U_\mu^\dagger (x-\hat{\mu}) q(x-\hat{\mu}) \Big] 
\ee
\be
\bar q (x) \stackrel{\leftarrow}{D_\mu} = \frac{1}{2} \, \Big[
\bar q (x+\hat{\mu}) U_\mu^\dagger (x) - \bar q(x-\hat{\mu}) 
U_\mu (x-\hat{\mu}) \Big] .
\ee

We have in some cases considered two representatives for an operator measuring
a given parton distribution. They are differentiated and identified by the 
choice of their Lorentz indices. The lattice operators corresponding to these 
choices fall in two different irreducible representations of the hypercubic 
group (the symmetry group of the lattice, the remnant of the Lorentz symmetry),
and on the lattice they will renormalize in a different way (whereas in a 
continuum scheme their renormalization factors are equal). In Monte Carlo 
measurements one of the two choices can be more convenient to use than the 
other, giving for instance smaller statistical and systematic errors, 
in particular when one considers the r\^ole played by nonvanishing momenta 
in numerical simulations.

Since we have done the calculations with $N_s=\infty$, an exact chiral 
symmetry is maintained in all our results, and its most important consequence 
is that the operator which measures the lowest moment of the $g_2$ structure 
function does not show any of the power-divergent mixings with operators of 
lower dimension which are instead present in the case of Wilson fermions. 
In fact, when chiral symmetry is broken $O_{d_1}$ mixes with a 
lower-dimensional operator which in the continuum operator product expansion 
is 
\be
m_q \, \bar q \gamma_{[4} \gamma_5 \gamma_{1]} q ,
\label{eq:g2mixing}
\ee 
but on the lattice instead has, in place of the mass, a $1/a$ coefficient 
which becomes infinite in the continuum limit. This mixing is forbidden 
for domain-wall fermions with infinite $N_s$, and $O_{d_1}$ is then in this
case multiplicatively renormalized. In addition, chiral symmetry implies that 
the renormalization constants of corresponding unpolarized and polarized 
operators (which differ by a $\gamma_5$ matrix) assume the same value. Thus, 
chiral symmetry gives a reduction of the number of independent renormalization 
factors in a given physical situation.

For operators which contain one covariant derivative one needs to perform 
a Taylor expansion of all vertices and propagators at first order in the 
lattice spacing $a$ (which means the external momentum $p$). We have chosen 
as loop integration momentum the one carried by the internal quarks. 
Choosing it the one carried by the gluon would result in much more complicated 
expressions for the order $p$ contributions.

\section{Results}
\label{sec:res}

\begin{figure}
\includegraphics[width=9.1cm]{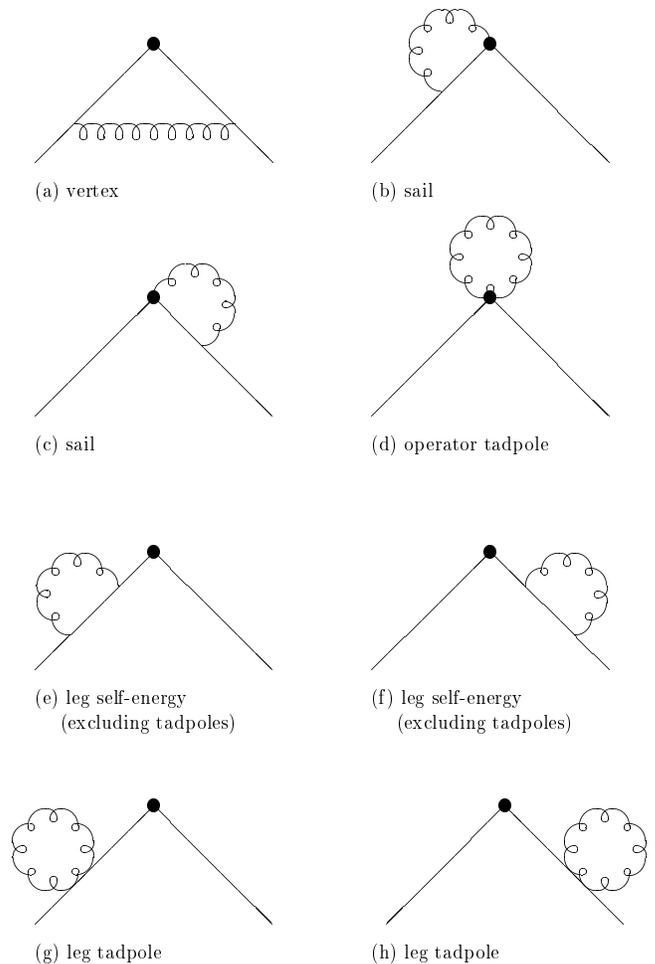}%
\caption{\label{fig:diagrams}The diagrams needed for the one-loop 
renormalization of the lattice operators.}
\end{figure}

\begin{table*}[htp]
\caption{\label{tab:results}
Values for the sums of the vertex and sail diagrams, $V^{\ \alpha=1}_O$, for 
the momentum, helicity and transversity operators considered in this work, 
in Feynman gauge. The one-loop results of the proper diagrams in a general 
covariant gauge can be inferred from Eqs.~(\ref{eq:properfirst}) to 
(\ref{eq:properlast}). We remind that 
$V^{\ \alpha=1}_{a_2,d} = V^{\ \alpha=1}_{v_2,d}$ and 
$V^{\ \alpha=1}_{a_2,e} = V^{\ \alpha=1}_{v_2,e}$.
The case of the tensor charge, $V^{\ \alpha=1}_{t_1}$, has been calculated 
for the first time in \cite{Aoki:1998vv}.}
\begin{ruledtabular}    
\begin{tabular}{|c|rrrrr|} 
$M$ & $V^{\ \alpha=1}_{v_2,d}$ & $V^{\ \alpha=1}_{v_2,e}$    
    & $V^{\ \alpha=1}_{d_1}$ & $V^{\ \alpha=1}_{t_1}$ & $V^{\ \alpha=1}_{t_2}$ 
      \vspace{0.05cm} \\ \hline \vspace{-0.3cm} \\  
    0.1    & -3.6205 & -3.2261 &  96.1427 & 5.1733 & -3.8636 \\
    0.2    & -3.5296 & -3.1111 &  42.4140 & 4.9150 & -3.8433 \\
    0.3    & -3.4553 & -3.0116 &  25.0326 & 4.7069 & -3.8223 \\
    0.4    & -3.3896 & -2.9194 &  16.5587 & 4.5245 & -3.8004 \\
    0.5    & -3.3288 & -2.8310 &  11.5737 & 4.3571 & -3.7776 \\
    0.6    & -3.2711 & -2.7441 &   8.2910 & 4.1987 & -3.7539 \\
    0.7    & -3.2150 & -2.6573 &   5.9508 & 4.0454 & -3.7291 \\
    0.8    & -3.1597 & -2.5694 &   4.1741 & 3.8943 & -3.7031 \\
    0.9    & -3.1042 & -2.4795 &   2.7486 & 3.7427 & -3.6758 \\
    1.0    & -3.0478 & -2.3865 &   1.5425 & 3.5882 & -3.6472 \\
    1.1    & -2.9898 & -2.2894 &   0.4655 & 3.4286 & -3.6170 \\
    1.2    & -2.9293 & -2.1869 &  -0.5533 & 3.2610 & -3.5850 \\
    1.3    & -2.8652 & -2.0776 &  -1.5787 & 3.0821 & -3.5510 \\
    1.4    & -2.7964 & -1.9598 &  -2.6834 & 2.8879 & -3.5148 \\
    1.5    & -2.7209 & -1.8311 &  -3.9675 & 2.6727 & -3.4760 \\
    1.6    & -2.6362 & -1.6880 &  -5.5992 & 2.4284 & -3.4342 \\
    1.7    & -2.5384 & -1.5258 &  -7.9262 & 2.1425 & -3.3889 \\
    1.8    & -2.4203 & -1.3360 & -11.8927 & 1.7930 & -3.3395 \\
    1.9    & -2.2671 & -1.1022 & -21.7691 & 1.3341 & -3.2850 \\ \hline
\end{tabular}
\end{ruledtabular}
\end{table*}

The diagrams required for the one loop lattice calculations of the matrix 
elements that we have considered here are given in Fig.~\ref{fig:diagrams}.
It can be easily seen that all tadpole diagrams are diagonal in the fifth 
dimension, and therefore they are equal to the expression calculated with
Wilson fermions. A leg tadpole has then the value 
\be
T_l = 8\pi^2 Z_0 \Big(1 - \frac{1}{4} (1-\alpha)\Big) ,
\ee
where $Z_0=0.154933390231\ldots$ is a well-known integral 
\cite{Capitani:2002mp}, while the operator tadpoles have the expression
\be
T_O = - T_l
\ee
for all operators considered in this work except the tensor charge, 
for which the operator tadpole vanishes, $T_{t_1} =0$  
\cite{Capitani:2000wi,Capitani:2000aq}.
The half-circle contribution of the quark self-energy as well as the vertex 
and sail diagrams instead have all a nontrivial structure in the fifth 
dimension, and due to their complexity we have deemed necessary to compute 
them using computer programs.

We have used the algebraic manipulation program FORM \cite{Vermaseren:2000nd}
to construct routines able to carry out all needed analytic calculations 
in an automated way. Sums of the damping factors in the fifth dimension 
and of the four-dimensional expressions in momentum space are evaluated 
with Fortran programs. To improve the convergence of the numerical integrals 
we use some of the techniques given by L\"uscher and Weisz 
in~\cite{Luscher:1985wf}.

The one-loop diagrams of Fig.~\ref{fig:diagrams} diverge at most 
logarithmically. Special care is required for the evaluation of the divergent 
terms, for which it is convenient to use the method introduced by Kawai 
{\it et al.} in~\cite{Kawai:1980ja} (see also \cite{Capitani:2002mp} for 
simple examples). A logarithmically divergent integral
\be
I(p) = \int dk \, \, {\cal I}(k,p) 
\ee 
is expanded in powers of the external momentum and split as
\be
I(p) = J(0) + (I(p)-J(0)) ,
\label{eq:divint}
\ee
where
\be
J(0) = \int \, dk \, \, {\cal I}(k,0) 
\ee
is its Taylor expansion to lowest order in $p$. Since the integrals appearing 
in $J$ do not depend on the external momentum, they are much easier to compute 
on the lattice than the complete integral of type $I$. The whole dependence 
on the external momentum remains in $I-J$, which is ultraviolet-finite for 
$a \rightarrow 0$ and can be evaluated by taking the naive continuum limit. 
Thanks to these facts, one is left with computing on the lattice only integrals
with vanishing momentum, which is technically straightforward. It is to be 
remarked that while $I$ is well defined, for finite lattice spacing both $J$ 
and $I-J$ are separately infrared divergent. To compute them one must then 
introduce an intermediate regularization, which we choose to be the naive 
dimensional regularization. The associated divergences will at the end cancel 
out in the sum $J + (I-J)$. 

To summarize, using this method the computation of any divergent integral 
which depends on an external momentum is reduced to the computation of  
lattice integrals at zero momentum plus some continuum integrals.

Identifying and processing divergent terms in an automated way for domain-wall 
fermions turns out to be somewhat more complicated and prone to errors than 
for simpler cases like Wilson fermions. We have thus devised an alternative 
indirect procedure for the evaluation of divergent integrals. This procedure 
uses the chain of equalities
\bea
I^{DW} &=& J^{DW} + (I^{DW} - J^{DW}) \\
       &=& J^{DW} + (I^{W} - J^{W})  \nonumber \\
       &=& (J^{DW} - J^{W}) + I^{W} ,
\eea
where $DW$ stands for domain-wall and $W$ for Wilson fermions.
In words, we numerically compute the difference between the domain-wall 
and Wilson zero-momentum $J$ integrals, and then add the full well-known 
Wilson result (for which several significant digits can be obtained 
without much effort). The key points here are that the difference 
$J^{DW} - J^{W}$ is a finite lattice integral, because the above-mentioned 
infrared divergences exactly cancel, and thus it does not need to be 
regularized at all, and moreover that the difference $I-J$ is an integral 
taken in the continuum limit, and so it makes no difference whether it is 
evaluated using domain-wall or Wilson fermions.

We have tested that our procedure works as desired by applying it to
calculations with overlap fermions, and we have reproduced in this way all 
results known in the literature for the bilinear operators 
\cite{Alexandrou:2000kj} and the first moment of the unpolarized parton 
distribution \cite{Capitani:2000wi,Capitani:2000aq}. This procedure is also 
much more precise than the subtraction of a known simple lattice integral 
with the same divergent behavior, which has the disadvantage of sometimes 
converging very slowly and hence it requires very large integration grids 
to attain the same accuracy. In a few cases we have used the simpler method
(which gives just a couple of significant digits) for consistency checks.

The amputated matrix elements that we have calculated have on the lattice 
the form (see Eq.~\ref{eq:matel1})
\be
1 + \bar g^2 \Big( - \gamma_O \log a^2 p^2 + B_O \Big) ,
\ee
with 
\be
B_O = V_O + T_O + \Sigma_1 ,
\ee
where $V_O$ is the finite contribution of the vertex and sail diagrams 
(a, b and c in Fig.~\ref{fig:diagrams}), $T_O$ refers to the tadpole arising 
from the operator (d in Fig.~\ref{fig:diagrams}), and $\Sigma_1$ is the finite 
contribution (proportional to $i\slash{p}$) from the quark self-energy of one 
leg, which also includes a leg tadpole (e and g, or f and h, in 
Fig.~\ref{fig:diagrams}). We call ``proper'' contributions the ones that do not
include the self-energy. They correspond to the diagrams a-d in 
Fig.~\ref{fig:diagrams}. Calling $1-\alpha = \xi $, the one-loop results 
for them are:
\bea
O_{v_2,d}^{proper} &=& \bar g^2 \, \Big( \Big( \frac{5}{3} 
+ \xi \Big) \log a^2 p^2  \label{eq:properfirst} \\ 
&& \quad + V^{\ \alpha=1}_{v_2,d} - \xi \cdot 6.850272  
+ T_{v_2,d} \Big) O_{v_2,d}^{tree} , \nonumber \\
O_{a_2,d}^{proper} &=& \bar g^2 \, \Big( \Big( \frac{5}{3} 
+ \xi \Big) \log a^2 p^2  \\ 
&& \quad + V^{\ \alpha=1}_{a_2,d} - \xi \cdot 6.850272 
+ T_{a_2,d} \Big) O_{a_2,d}^{tree} , \nonumber \\
O_{v_2,e}^{proper} &=& \bar g^2 \, \Big( \Big( \frac{5}{3} 
+ \xi \Big) \log a^2 p^2 \\ 
&& \quad + V^{\ \alpha=1}_{v_2,e} - \xi \cdot 6.850272
+ T_{v_2,e} \Big) O_{v_2,e}^{tree} , \nonumber \\
O_{a_2,e}^{proper} &=& \bar g^2 \, \Big( \Big( \frac{5}{3} 
+ \xi \Big) \log a^2 p^2 \\ 
&& \quad + V^{\ \alpha=1}_{a_2,e} - \xi \cdot 6.850272
+ T_{a_2,e} \Big) O_{a_2,e}^{tree} , \nonumber \\
O_{d_1}^{proper} &=& \bar g^2 \, \Big( (-1 + \xi ) \log a^2 p^2  \\
&& \quad +V^{\ \alpha=1}_{d_1} - \xi \cdot 7.850272
+ T_{d_1} \Big) O_{d_1}^{tree} , \nonumber  \\
O_{t_1}^{proper} &=& \bar g^2 \, \Big( 
\xi \log a^2 p^2  \\
&& \quad +V^{\ \alpha=1}_{t_1} - \xi \cdot 3.792010
\Big) O_{t_1}^{tree} , \nonumber \\
O_{t_2}^{proper} &=& \bar g^2 \, \Big( \Big( 2 
+ \xi \Big) \log a^2 p^2  \label{eq:properlast} \\
&& \quad +V^{\ \alpha=1}_{t_2} - \xi \cdot 6.350272
+ T_{t_2} \Big) O_{t_2}^{tree} \nonumber .
\eea
The results due to the sails and vertices, $V_O$, have been for convenience
separated in the Feynman gauge values $V^{\ \alpha=1}_O$, listed in Table 
\ref{tab:results} for various choices of the Dirac mass $M$ between $M=0.1$
and $M=1.9$, and the remaining contributions proportional to $\xi$, 
which are instead independent of $M$ and shown in the above equations.
Notice that also the tadpoles coming from the operators provide contributions
proportional to $\xi$. The results for the tensor charge, $V_{t_1}$, are equal 
to the results for the standard tensor current $\bar q \sigma_{\mu\nu} q$, 
which were already calculated in \cite{Aoki:1998vv}, and are reported, together
with the other bilinears and the self-energy results, in the Appendix.

A significant check of our perturbative calculations is that the contributions
proportional to $\xi$ in covariant gauge are constant in $M$, as already noted 
in \cite{Capitani:2000wi,Capitani:2000aq,Capitani:2002mp} for the case of 
overlap fermions and discussed more in depth in \cite{Horsley:2004mx}.
Furthermore, they are equal to the results obtained with Wilson fermions 
(and this is the reason why we can provide more significant digits for these 
contributions). They are also independent of the lattice representation of the 
operator (e.g., for $O_{v_2,d}$ and $O_{v_2,e}$). Their analytic expressions 
are very complicated and highly nonlinear functions of $M$ containing hundreds
of terms, and the numerical cancellation of this dependence is a rather strong 
check on the good behavior of the FORM codes, as well as of the integration 
routines. 

Another reasonably strong check is that the operators $O_{v_2,d}$ and 
$O_{a_2,d}$ have the same renormalization constant well within the numerical 
integration errors, as expected from chiral symmetry. We have checked that 
this is also true for the pair $O_{v_2,e}$ and $O_{a_2,e}$. Thus, in all cases 
the polarized operators have the same renormalization constants as the 
corresponding unpolarized operators. Furthermore, as we have explicitly 
verified, for the same reason the $1/a$ coefficient of the mixing term 
of Eq.~(\ref{eq:g2mixing}) arising in the one-loop expression of the operator 
$O_{d_1}$ tends to zero when the integration grid is refined. This operator 
is then for domain-wall fermions multiplicatively renormalized, contrary to 
what happens in the Wilson case, where its mixing coefficient goes to infinity 
in the continuum limit. 

In the numerical integration, the convergence can become slow when $M$ is 
very close to zero or two. Thus, while otherwise a grid of 60 or 80 points 
in each direction is sufficient to obtain about five significant digits, for 
$M=0.1$ and $M=1.9$ we had sometimes to increase the grid to 100 points in 
each direction in order to achieve the same precision. 

\begin{table*}[htp]
\caption{\label{tab:zetas}
Values of the renormalization factors for the various operators, for 
domain-wall QCD at $g_0=1$ and $\mu=1/a$. We remind that 
$Z_{a_2,d} = Z_{v_2,d}$ and $Z_{a_2,e} = Z_{v_2,e}$.}
\begin{ruledtabular}    
\begin{tabular}{|c|rrrrr|} 
$M$ & $Z_{v_2,d}$ & $Z_{v_2,e}$    
    & $Z_{d_1}$ & $Z_{t_1}$ & $Z_{t_2}$ 
      \vspace{0.05cm} \\ \hline \vspace{-0.3cm} \\  
  0.1  & 0.9979 & 0.9945 & 0.1931 & 0.8494 & 0.9952 \\     
  0.2  & 0.9984 & 0.9948 & 0.6480 & 0.8529 & 0.9963 \\     
  0.3  & 0.9988 & 0.9951 & 0.7958 & 0.8557 & 0.9972 \\     
  0.4  & 0.9992 & 0.9952 & 0.8683 & 0.8582 & 0.9980 \\     
  0.5  & 0.9995 & 0.9953 & 0.9112 & 0.8604 & 0.9986 \\     
  0.6  & 0.9997 & 0.9953 & 0.9396 & 0.8624 & 0.9991 \\     
  0.7  & 0.9998 & 0.9951 & 0.9600 & 0.8643 & 0.9995 \\     
  0.8  & 0.9999 & 0.9949 & 0.9755 & 0.8661 & 0.9998 \\     
  0.9  & 0.9998 & 0.9946 & 0.9879 & 0.8678 & 1.0000 \\     
  1.0  & 0.9997 & 0.9941 & 0.9984 & 0.8694 & 1.0000 \\     
  1.1  & 0.9994 & 0.9935 & 1.0077 & 0.8710 & 1.0000 \\     
  1.2  & 0.9990 & 0.9927 & 1.0164 & 0.8725 & 0.9998 \\     
  1.3  & 0.9984 & 0.9917 & 1.0250 & 0.8740 & 0.9995 \\     
  1.4  & 0.9976 & 0.9905 & 1.0342 & 0.8754 & 0.9990 \\     
  1.5  & 0.9966 & 0.9890 & 1.0446 & 0.8768 & 0.9982 \\     
  1.6  & 0.9952 & 0.9872 & 1.0578 & 0.8782 & 0.9973 \\     
  1.7  & 0.9934 & 0.9849 & 1.0765 & 0.8797 & 0.9959 \\     
  1.8  & 0.9911 & 0.9819 & 1.1086 & 0.8813 & 0.9942 \\     
  1.9  & 0.9878 & 0.9780 & 1.1900 & 0.8832 & 0.9917 \\ \hline    
\end{tabular}
\end{ruledtabular}
\end{table*}

To obtain the complete one-loop amplitudes we have now to add to the results
of the proper diagrams the 1-loop contributions of the self-energy which are 
proportional to $i\slash{p}$,
\be
\Sigma_1 = \bar g^2 \Big( (1 - \xi) \log a^2 p^2  
+\Sigma_1^{\ \alpha=1} + \xi \cdot 4.792010 \Big),
\label{eq:selfenergy}
\ee
where $\Sigma_1^{\ \alpha=1}=10.8750$ when $M=1$, while for other values 
of $M$ the Feynman-gauge finite terms $\Sigma_1^{\ \alpha=1}$ are given in 
the Appendix. The complete one-loop lattice results are then, for $M=1$:
\bea
O_{v_2,d}^{lat} \! &=& \! \Big( 1 + \bar g^2 
\Big(\frac{8}{3} \log a^2 p^2 -4.4059 +\xi \Big) \Big) O_{v_2,d}^{tree} , \\
O_{a_2,d}^{lat} \! &=& \! \Big( 1 + \bar g^2 
\Big(\frac{8}{3} \log a^2 p^2 -4.4059  +\xi \Big) \Big) O_{a_2,d}^{tree} , \\
O_{v_2,e}^{lat} \! &=& \! \Big( 1 + \bar g^2 
\Big(\frac{8}{3} \log a^2 p^2 -3.7445 +\xi \Big) \Big) O_{v_2,e}^{tree} , \\
O_{a_2,e}^{lat} \! &=& \! \Big( 1 + \bar g^2  
\Big(\frac{8}{3} \log a^2 p^2 -3.7445  +\xi \Big) \Big) O_{a_2,e}^{tree} , \\
O_{d_1}^{lat} \! &=& \! \Big( 1 + \bar g^2 \cdot 0.1845 \Big) 
O_{d_1}^{tree} , \qquad \qquad \qquad \qquad \\
O_{t_1}^{lat} \! &=& \! \Big( 1 + \bar g^2 
\Big(\log a^2 p^2 +14.4633 + \xi \Big) \Big) O_{t_1}^{tree} , \\
O_{t_2}^{lat} \! &=& \! \Big( 1 + \bar g^2 
\Big(3 \log a^2 p^2 -5.0052 + \frac{3}{2} \, \xi \Big) \Big) O_{t_2}^{tree} . 
\eea
To establish the connection with the corresponding continuum quantities we also
need to know the one-loop amplitudes for the same operators in the $\ms$ scheme
\cite{Capitani:2000wi,Capitani:2000aq}:
\bea
O^{\ms}_{v_2} 
&=& \Big( 1 + \bar g^2 \Big( \frac{8}{3} 
\log \frac{p^2}{\mu^2} -\frac{40}{9} +\xi \Big) \Big) 
O_{v_2}^{tree} , \\ 
O^{\ms}_{a_2} 
&=& \Big( 1 + \bar g^2 \Big( \frac{8}{3} 
\log \frac{p^2}{\mu^2} -\frac{40}{9} +\xi \Big) \Big) 
O_{a_2}^{tree} , \\ 
O^{\ms}_{d_1} 
&=& O_{d_1}^{tree} , \\ 
O^{\ms}_{t_1} 
&=&\Big( 1 + \bar g^2 \Big(\log \frac{p^2}{\mu^2}  
-1 +\xi \Big) \Big) O_{t_1}^{tree} , \\ 
O^{\ms}_{t_2} 
&=&\Big( 1 + \bar g^2 \Big(3 \log \frac{p^2}{\mu^2} 
-5 +\frac{3}{2} \, \xi \Big) \Big) O_{t_2}^{tree} .
\eea
Putting all together, we obtain the factors that allow the matching from the 
domain-wall lattice theory to the $\ms$ continuum scheme, for $M=1$:
\bea
O^{\ms}_{v_2,d} 
&=&\Big( 1 - \bar g^2 \Big(\frac{8}{3} \log a^2 \mu^2 
+0.0386 \Big) \Big) O^{lat}_{v_2,d} , \\ 
O^{\ms}_{a_2,d} 
&=&\Big( 1 - \bar g^2 \Big(\frac{8}{3} \log a^2 \mu^2 
+0.0386  \Big) \Big) O^{lat}_{a_2,d} , \\ 
O^{\ms}_{v_2,e} 
&=&\Big( 1 - \bar g^2 \Big(\frac{8}{3} \log a^2 \mu^2 
+0.6999 \Big) \Big) O^{lat}_{v_2,e} , \\ 
O^{\ms}_{a_2,e} 
&=&\Big( 1 - \bar g^2 \Big(\frac{8}{3} \log a^2 \mu^2 
+0.6999  \Big) \Big) O^{lat}_{a_2,e} , \\ 
O^{\ms}_{d_1} 
&=&\Big( 1 + \bar g^2 \cdot 0.1845 \Big) 
O^{lat}_{d_1} , \\ 
O^{\ms}_{t_1} 
&=&\Big( 1 - \bar g^2 \Big(\log a^2 \mu^2 
+15.4633 \Big) \Big) O^{lat}_{t_1} , \\ 
O^{\ms}_{t_2} 
&=&\Big( 1 - \bar g^2 \Big( 3 \log a^2 \mu^2 
-0.0052 \Big) \Big) O^{lat}_{t_2} . 
\eea
Notice that the part proportional to $\xi$ has canceled between the lattice 
and continuum expressions, and these renormalization factors are hence gauge
invariant.

For simulations of domain-wall QCD at $g_0=1$, setting $\mu=1/a$ one obtains 
the values
\bea
O^{\ms}_{v_2,d} 
&=& 0.9997 \cdot O^{lat~(domain-wall,~M=1.0)}_{v_2,d} \\
&=& 0.98920 \cdot O^{lat~(Wilson)}_{v_2,d} , \nonumber \\ 
O^{\ms}_{a_2,d} 
&=& 0.9997 \cdot O^{lat~(domain-wall,~M=1.0)}_{a_2,d} \\ 
&=& 0.99709 \cdot O^{lat~(Wilson)}_{a_2,d} , \nonumber \\ 
O^{\ms}_{v_2,e} 
&=& 0.9941 \cdot O^{lat~(domain-wall,~M=1.0)}_{v_2,e} \\
&=& 0.97837 \cdot O^{lat~(Wilson)}_{v_2,e} , \nonumber \\ 
O^{\ms}_{a_2,e} 
&=& 0.9941 \cdot O^{lat~(domain-wall,~M=1.0)}_{a_2,e} \\ 
&=& 0.99859 \cdot O^{lat~(Wilson)}_{a_2,e} , \nonumber \\ 
O^{\ms}_{d_1} 
&=& 0.9984 \cdot O^{lat~(domain-wall,~M=1.0)}_{d_1} , \\ 
O^{\ms}_{t_1} 
&=& 0.8694 \cdot O^{lat~(domain-wall,~M=1.0)}_{t_1} \\  
&=& 0.85631 \cdot O^{lat~(Wilson)}_{t_1} , \nonumber \\ 
O^{\ms}_{t_2} 
&=& 1.0000 \cdot O^{lat~(domain-wall,~M=1.0)}_{t_2} \\
&=& 0.99559 \cdot O^{lat~(Wilson)}_{t_2} , \nonumber 
\eea
where for comparison the corresponding Wilson results are also shown. 
Of course the domain-wall renormalization factors vary with $M$. 
For example, for $M=1.8$ (which is almost at the edge of the allowed 
parameter space) their values are instead
\bea
O^{\ms}_{v_2,d} 
&=& 0.9911 \cdot O^{lat~(domain-wall,~M=1.8)}_{v_2,d} , \\
O^{\ms}_{a_2,d} 
&=& 0.9911  \cdot O^{lat~(domain-wall,~M=1.8)}_{a_2,d} , \\ 
O^{\ms}_{v_2,e} 
&=& 0.9819 \cdot O^{lat~(domain-wall,~M=1.8)}_{v_2,e} , \\
O^{\ms}_{a_2,e} 
&=& 0.9819 \cdot O^{lat~(domain-wall,~M=1.8)}_{a_2,e} , \\ 
O^{\ms}_{d_1} 
&=& 1.1086 \cdot O^{lat~(domain-wall,~M=1.8)}_{d_1} , \\ 
O^{\ms}_{t_1} 
&=& 0.8813 \cdot O^{lat~(domain-wall,~M=1.8)}_{t_1} , \\  
O^{\ms}_{t_2} 
&=& 0.9942 \cdot O^{lat~(domain-wall,~M=1.8)}_{t_2} . 
\eea
Results for other choices of $M$ and $g_0$ can be easily obtained from the 
numbers presented in this Section. Table \ref{tab:zetas} contains the
values of the renormalization factors as a function of $M$ for $g_0=1$.

All the renormalization constants presented in this work can also be used
in unquenched simulations, provided that one computes only matrix elements 
of flavor nonsinglet quark operators, for which at one-loop internal quark 
loops never appear. The numbers for the transversity operators can be however 
considered unquenched even in the singlet case, since no chiral-odd gluon 
operators exist, which would constitute the only possibility for having 
a mixing.

We can easily notice that the renormalization corrections that we have obtained
for domain-wall fermions are in general small. In particular when the Dirac 
mass is $M=1$ or not too far from it they are not too different from the 
corresponding Wilson results. In many cases the total renormalization factors 
turn out to be quite close to one, reflecting the fact that the domain-wall 
one-loop amplitudes have almost the same values as the corresponding $\ms$ 
results. The only exception is the tensor charge, which on the other hand is 
also the only case which for fermions which break chiral symmetry cannot be 
renormalized, because of its power-divergent mixing. From this point of view,
domain-wall fermions appear to behave quite at variance with overlap fermions, 
for which the renormalization factors are generally not small, giving in many 
cases rather large one-loop corrections to the tree-level matrix elements 
\cite{Alexandrou:2000kj,Capitani:2000wi,Capitani:2000aq}. 
The origin of most of these large effects can be traced back to the $\Sigma_1$ 
contribution from the half-circle diagram of the self-energy, which for overlap
fermions is rather big. On the contrary, for domain-wall fermions $\Sigma_1$ 
does not deviate too much from the $\Sigma_1$ of Wilson fermions. Thus, apart 
from $O_{d_1}$ when $M$ is away from one, the renormalization factors computed
in this work give small corrections at the standard accessible couplings.

We have computed the bilinears and the self-energy anew, and we have found 
some discrepancies when comparing our results, which we report in the Appendix,
with the numbers given in \cite{Aoki:1998vv}. These discrepancies derive 
only from the continuum integrals that are needed to compute the divergent 
terms with the Kawai method (for $\Sigma_3$ and the tensor charge, which are 
finite, we completely agree). The differences between our results for the 
$R^{lat}$ quantities and those in \cite{Aoki:1998vv} are indeed in all cases 
independent of $M$ and always are an integer or half-integer number. 
The $\ms$ renormalization factors given in \cite{Aoki:1998vv} are also 
different, and for the same amount, from what is found elsewhere in the 
literature (e.g., \cite{Alexandrou:2000kj,Capitani:2000wi,Capitani:2000aq}). 
All these differences cancel then in the expressions of the renormalization 
factors, which do not present any discrepancies with \cite{Aoki:1998vv}.

\section{Conclusions}
\label{sec:concl}

In this paper we have presented the computation of the one-loop renormalization
factors of a few operators which measure the first nontrivial moment of 
various structure functions, giving a complete description of the quark 
momentum and spin at leading twist. We have used domain-wall fermions, and 
the associated chiral symmetry plays an important r\^ole in the structure of 
the strong radiative corrections.

We have automated the calculations by developing suitable FORM codes.
This will make it easier to consider the case of more complicated operators.
The renormalization factors that we have found turn out to be in many cases 
close to one.

\begin{acknowledgments}
I am grateful for the support by Fonds zur F\"orderung der Wissenschaftlichen 
Forschung in \"Osterreich (FWF), Project P16310-N08.
\end{acknowledgments}

\appendix*

\section{Self-energy and bilinears}

\begin{table*}[htp]
\caption{\label{tab:bilinears}
Values of the domain-wall constants needed for the renormalization of the 
self-energy and the bilinear operators. We remind that 
$V^{\ \alpha=1}_P = V^{\ \alpha=1}_S$ and 
$V^{\ \alpha=1}_A = V^{\ \alpha=1}_V$, and that 
$z_w = - 2w_0 \, \Sigma_3/(1-w_0^2)$.}
\begin{ruledtabular}    
\begin{tabular}{|c|rrrrr|} 
$M$ & $\Sigma_1^{\ \alpha=1}$   & $\Sigma_3$   & $z_w$ 
    & $V_S^{\ \alpha=1}$   & $V_V^{\ \alpha=1}$
      \vspace{0.05cm} \\ \hline \vspace{-0.3cm} \\  
    0.1    & 11.6603 & 51.0482 & -483.6145 &  7.8219 & 6.3355 \\
    0.2    & 11.5099 & 50.7450 & -225.5333 &  8.6070 & 6.3380 \\
    0.3    & 11.3829 & 50.4885 & -138.5959 &  9.2424 & 6.3408 \\
    0.4    & 11.2730 & 50.2664 &  -94.2495 &  9.8020 & 6.3438 \\
    0.5    & 11.1772 & 50.0726 &  -66.7635 & 10.3176 & 6.3472 \\
    0.6    & 11.0939 & 49.9038 &  -47.5274 & 10.8074 & 6.3509 \\
    0.7    & 11.0221 & 49.7582 &  -32.8076 & 11.2834 & 6.3549 \\
    0.8    & 10.9616 & 49.6352 &  -20.6813 & 11.7549 & 6.3594 \\
    0.9    & 10.9124 & 49.5351 &  -10.0071 & 12.2296 & 6.3644 \\
    1.0    & 10.8750 & 49.4588 &    0.0000 & 12.7151 & 6.3699 \\
    1.1    & 10.8504 & 49.4084 &    9.9815 & 13.2189 & 6.3762 \\
    1.2    & 10.8399 & 49.3865 &   20.5777 & 13.7496 & 6.3831 \\
    1.3    & 10.8455 & 49.3972 &   32.5696 & 14.3176 & 6.3910 \\
    1.4    & 10.8699 & 49.4461 &   47.0915 & 14.9360 & 6.4000 \\
    1.5    & 10.9170 & 49.5411 &   66.0549 & 15.6225 & 6.4102 \\
    1.6    & 10.9923 & 49.6935 &   93.1753 & 16.4024 & 6.4219 \\
    1.7    & 11.1041 & 49.9198 &  137.0349 & 17.3150 & 6.4356 \\
    1.8    & 11.2652 & 50.2462 &  223.3164 & 18.4274 & 6.4516 \\
    1.9    & 11.4979 & 50.7176 &  480.4824 & 19.8800 & 6.4706 \\ 
\hline
\end{tabular}
\end{ruledtabular}
\end{table*}

We report here the results for the quark self-energy and the bilinear 
operators, which were first calculated in \cite{Aoki:1998vv} in Feynman gauge. 

Taking into account that the results for the pseudoscalar and axial-vector
operators are equal to the ones for the scalar and vector operators 
respectively, and that the tensor operator has been reported in the main body 
of the paper, the one-loop results for the proper diagrams that we need here 
are:
\pagebreak
\bea
O_S^{proper} &=& \bar g^2 \, \Big( ( -4 + \xi ) \log a^2 p^2 \\
&& \quad + V^{\ \alpha=1}_S - \xi \cdot 5.792010
\Big) O_S^{tree} , \nonumber \\
O_V^{proper} &=& \bar g^2 \, \Big( ( -1 + \xi ) \log a^2 p^2 \\ 
&& \quad + V^{\ \alpha=1}_V - \xi \cdot 4.792010
\Big) O_V^{tree} . \nonumber 
\eea
There is no operator tadpole for the bilinears, and adding the self-energy 
contribution proportional to $i\slash{p}$ (Eq.~(\ref{eq:selfenergy}))
\be
\Sigma_1 = \bar g^2 \Big( (1 - \xi) \log a^2 p^2  
+\Sigma_1^{\ \alpha=1} + \xi \cdot 4.792010 \Big),
\ee
we get, for $M=1$,
\bea
O_S^{lat} \! &=& \! \Big( 1 - \bar g^2 \Big( 3 \log a^2 p^2 
-23.5901 + \xi \Big) \Big) O_S^{tree} , \\
O_V^{lat} \! &=& \! \Big( 1 + \bar g^2 \cdot 17.2450 \Big) O_V^{tree} .
\eea
The one-loop results in the $\ms$ scheme are 
\cite{Alexandrou:2000kj,Capitani:2000wi,Capitani:2000aq}:
\bea
O^{\ms}_S 
&=& \Big( 1 + \bar g^2 \Big( -3
\log \frac{p^2}{\mu^2} +5 - \xi \Big) \Big) O_S^{tree} , \\ 
O^{\ms}_V &=& O_V^{tree} , 
\eea
and thus
\bea
O^{\ms}_S 
&=& 0.8430 \cdot O^{lat~(domain-wall,~M=1.0)}_S \\
&=& 0.89064 \cdot O^{lat~(Wilson)}_S , \nonumber \\ 
O^{\ms}_P
&=& 0.8430 \cdot O^{lat~(domain-wall,~M=1.0)}_P \\ 
&=& 0.80922 \cdot O^{lat~(Wilson)}_P , \nonumber \\ 
O^{\ms}_V 
&=& 0.8544 \cdot O^{lat~(domain-wall,~M=1.0)}_V \\
&=& 0.82592 \cdot O^{lat~(Wilson)}_V , \nonumber \\ 
O^{\ms}_A 
&=& 0.8544 \cdot O^{lat~(domain-wall,~M=1.0)}_A \\ 
&=& 0.86663 \cdot O^{lat~(Wilson)}_A , \nonumber  
\eea
where for comparison the corresponding Wilson results are also shown. 
The renormalization factors for other values of $M$ and $g_0$ can be easily 
obtained from the numbers given in Table \ref{tab:bilinears}, where we 
in addition to $\Sigma_1$ we report also the results for the quantities 
$\Sigma_3$ and
\be
z_w = - \frac{2w_0}{1-w_0^2} \, \Sigma_3 ,
\ee
which determines the amount of additive renormalization to $w_0=1-M$.

\end{document}